\documentclass[twocolumn,english,aps]{revtex4}
\usepackage[T1]{fontenc}
\usepackage[latin9]{inputenc}
\usepackage{graphicx}
\usepackage{amssymb}

\makeatletter

\usepackage{babel}
\makeatother

\begin{document}

\title{Spin dynamics of S-state ions in the filled skutterudites La$_{1-x}$$R$$_{x}$Pt$_{4}$Ge$_{12}$
($R=$ Gd, Eu) }

\author{F.A. Garcia, R. Gumeniuk, W. Schnelle, J. Sichelschmidt, A. Leithe-Jasper,
Yu. Grin, F. Steglich }

\address{Max-Planck-Institut fur Chemische Physik fester Stoffe, Nothnitzer
Strabe 40, 01187 Dresden, Germany.}

\date{02/09/12}

\begin{abstract}
A detailed study of the spin dynamics of the S-state ions Gd$^{3+}$
and Eu$^{2+}$ in the filled skutterudites La$_{1-x}$$R$$_{x}$Pt$_{4}$Ge$_{12}$
($R=$ Gd, Eu) is reported. The spin dynamics is investigated directly
by means of Gd$^{3+}$ and Eu$^{2+}$electron spin resonance (ESR),
performed at $X$-band ($\approx9.4$ GHz) and $Q$-band ($\approx34$
GHz) frequencies in the temperature intervals $8<T<300$ K and $1.5<T<300$
K respectively. The ESR parameters provide direct evidence for the
vibrational behavior of the Gd$^{3+}$ ions but not for the Eu$^{2+}$
ions. These results are interpreted in the light of recent discussions
about the spin relaxation in cage systems. In particular, the Gd$^{3+}$
spin relaxation in La$_{0.9}$Gd$_{0.1}$Pt$_{4}$Ge$_{12}$ provide
evidence for the existence of an extra phonon mode with an Einstein
temperature $\theta_{E}\approx24$ K in this system. The work suggests
that the so-called {}``rattling'' modes have a general and important
role that should be taken into account in the study of spin dynamics
in cage systems. 
\end{abstract}
\maketitle

\section{Introduction}

Several families of intermetallic compounds are today known to crystallize
in cage-like structures. A topic of growing interest in this broad
area of research is the role of the lattice dynamics in the physical
properties found in these materials. Among these cage systems, the
family of the filled skutterudites has been attracting special attention
triggered by the wide range of physical phenomena found in these compounds.
These include exotic strongly correlated ground states \citep{maple_new_2009}
as well as a relatively large Seebeck coefficient, which opens the
perspective of using skutterudites in the construction of thermoelectric
devices \citep{snyder_complex_2008}. 

Filled skutterudites have the general formula R$_{y}$T$_{4}$X$_{12}$
and crystallize in the LaFe$_{4}$P$_{12}$ structure with space group
$Im\overline{3}$ and local point symmetry T$_{h}$ for the R ions
\citep{jeitschko_lafe4p12_1977}. The R element is usually referred
as \emph{guest,} or \emph{filler,} and resides in the large void in
the framework of the {[}T$_{4}$X$_{12}$] \emph{host} structure.
Systematic analysis of the chemical bonding in skutterudite compounds
\citep{leithe-jasper_weak_2004,gumeniuk_filled_2010} revealed a fundamental
role of the R ions in the stabilization of the crystal structure.

It is generally accepted that in many of these compounds, a substantial
change in the collective vibrational dynamics of the system is implied
by the dynamical behavior of the R ions \citep{koza_vibrational_2010,koza_vibrational_2011}.
In the simplest approximation, the R-ion dynamics is described in
terms of a localized and isolated phonon mode, usually called {}``rattling''
mode, and it can be fully described by a single parameter $\theta_{E}$,
the Einstein temperature \citep{keppens_localized_1998,cao_evidence_2004,yamaura_rattling_2011}.
However, the nature and role of the total vibrational dynamics has
been a matter of intense debate in the field, and there are a growing
number of experimental \citep{koza_breakdown_2008,koza_vibrational_2010,schnelle_magnetic_2008,goto_quadrupolar_2004,tsubota_smsb_2008,koza_vibrational_2011}
and theoretical \citep{hattori_local_2005,yashiki_kondo_2011,oshiba_strong-coupling_2011}
investigations demonstrating the inadequacy, and exploring possible
extensions, of this scenario. Of special interest for the present
work are current studies of the spin dynamics in skutterudites, and
also in other cage systems, by means of electron spin resonance (ESR)
and also nuclear magnetic resonance (NMR) \citep{sichelschmidt_electron_2005,nakai_low-lying_2008,holanda_electron_2009,nowak_139la_2011,garcia_eu2+_2011,kanetake_superconducting_2010,baenitz_ge-based_2010,magishi_nmr_2010}. 

A recent theoretical work by Dahm \emph{et al.} \citep{dahm_nmr_2007}
investigated the NMR relaxation from rattling modes, including anharmonic
ones. It was found that, even in metallic skutterudites, the rattling
behavior may give rise to an important contribution to the spin relaxation
by means of a Raman process (two-phonon coupling) which would coexist
with the usual Korringa process (conduction-electron scattering).
Recent NMR experiments in the skutterudites LaOs$_{4}$Sb$_{12}$
\citep{nakai_low-lying_2008} and LaPt$_{4}$Ge$_{12}$ \citep{kanetake_superconducting_2010}
have successfully identified an additional contribution to the Korringa
relaxation, that was analyzed in the framework of the above cited
theory \citep{dahm_nmr_2007}. The relaxation at the La site was shown
to be field independent, which was interpreted as a piece of evidence
for a non-magnetic origin of this additional relaxation. 

In a recent work, the spin dynamics of Eu$^{2+}$ in the EuT$_{4}$Sb$_{12}$
(T = Fe, Ru, Os) skutterudites was investigated by ESR \citep{garcia_eu2+_2011}
and clear signatures of the Eu$^{2+}$ dynamical behavior were observed.
It was found that the ESR linewidth $\Delta H$, as a function of
temperature, peaks at about $\theta_{E}$ and increases linearly at
higher temperatures. Nevertheless, in contrast with the NMR experiments,
the ESR results were shown to be field dependent. Thus, the results
were ascribed to inhomogeneities of the spin-spin interaction, implied
by the rattling modes. It was suggested that these inhomogeneities
are quenched at high fields, causing the above features in $\Delta H$
to disappear. 

A fundamental issue in the application of ESR to probe the interplay
of localized spins and the guest-ion dynamics in metallic skutterudites
is the bottleneck effect \citep{barnes_theory_1981}. The bottleneck
effect in metals occurs when the conduction electron ($\mathrm{ce}$)-lattice
($\mathrm{L}$) relaxation time ($1/T_{\mathrm{ceL}}$) is slow in
comparison with the conduction electron-localized spin ($\mathrm{S}$)
relaxation time $(1/T_{\mathrm{ceS}})$. Hence, after receiving energy
from the localized spins, the conduction electrons instead of dissipating
energy to the lattice, give this energy back to the localized spins.
As a consequence, the relaxation of the localized spins is now modulated
by the slow $1/T_{\mathrm{ceL}}$. In turn, one should always be careful
in drawing any conclusions about a relation between effects in $\Delta H$
and effects in the exchange interaction between the local and itinerant
spins.

Aiming to discuss the applicability of the ESR technique in the study
of the spin dynamics in cage systems, here, we present at two microwave
frequencies ($X$- and $Q$-band) the ESR spectra of Gd$^{3+}$ and
Eu$^{2+}$ (configuration $J=S=7/2$, $L=0$) in La$_{1-x}$$R$$_{x}$Pt$_{4}$Ge$_{12}$
($R=$ Gd, Eu) for various values of $x$. Choosing S-state ions,
we rule out non-trivial crystal-field ($\mathrm{CF}$) effects implied
by the local $T_{h}$ symmetry of the {}``filler'' site \citep{takegahara_crystal_2001}.
In addition, we address the difference between the substitution of
La$^{3+}$ by small amounts of Gd$^{3+}$ (isoelectronic substitution)
and Eu$^{2+}$ (hole doping). 

The filled skutterudite LaPt$_{4}$Ge$_{12}$ is a metallic compound
which undergoes a superconducting transition at $T_{\mathrm{sc}}=8.3$
K \citep{gumeniuk_superconductivity_2008}. Electronic structure calculations
show that the Fermi surface of this compound is mainly composed by
$4p$ orbitals from the Ge atoms, with rather small contributions
derived from the Pt $5d$ orbitals. Magnetization measurements show
no signs for magnetic fluctuations or itinerant moments. The superconducting
state, however, is not yet fully understood \citep{maisuradze_evidence_2010,toda_electrical_2008}
and we address the partial suppression of the superconducting state
due to the inclusion of Gd$^{3+}$ moments. 

The paper is organized as follows: first, we present and discuss our
experiments with the Gd doped samples. Then, we contrast these results
with the experiments on the Eu-doped sample and also with EuPt$_{4}$Ge$_{12}$.
Afterward, we explore our results in the context of relaxation from
rattling modes \citep{dahm_nmr_2007}.

\section{Experiment}

Single-phase polycrystalline samples of La$_{1-x}$Gd$_{x}$Pt$_{4}$Ge$_{12}$
($x=0.005$, $0.01$, $0.05$ and $0.1$) and La$_{1-x}$Eu$_{x}$Pt$_{4}$Ge$_{12}$
($x=0.01$, $0.05$, $1$) were prepared as described elsewhere \citep{gumeniuk_superconductivity_2008,gumeniuk_filled_2010}.
To verify the content of the S-state ions, the temperature dependence
of the DC susceptibility of each sample was measured and the experimental
results were fitted to a Curie-Weiss behavior, assuming the full effective
moment $\mu_{\mathrm{eff}}=7.93$ $\mu_{B}$ for both Gd$^{3+}$ and
Eu$^{2+}$ ions. These measurements were performed in a commercial
SQUID magnetometer (Quantum Design). The ESR measurements were performed
in a Bruker Elexsys 500 spectrometer at both $X$-band ($\nu=9.4$
GHz) and $Q$-band ($\nu=34.4$ GHz) frequencies, in the temperatures
intervals $8<T<300$ K and $1.5<T<300$ K respectively. The $Q$-band
measurements at relatively lower temperatures were possible because
the typical applied magnetic fields close to the resonance field $H_{res}$
at $Q$-band ($H_{res}\approx1.2$ T, see below) used in the ESR measurements
are larger than the upper critical field $H_{c2}$ ( $H_{c2}\approx1$
T at $T\approx7$ K ) measured for LaPt$_{4}$Ge$_{12}$ \citep{gumeniuk_filled_2010}.

The samples where powdered and sieved (gritting of $40$$\mu$m) for
grain size homogeneity. The size of these grains is still much larger
than the skin depth that can be estimated from the resistivity measurements
in the temperature interval of the experiment. Hence, we observed
an asymmetric resonance line that was fitted as described in Ref.
\citep{joshi_analysis_2004}. The parameters included in the fitting
are the ESR linewidth $\Delta H$, the resonance field $H_{res}$,
which defines the ESR $g$-values by applying the resonance condition
$\hbar\nu=g\mu_{B}H_{res}$, the resonance amplitude and an $\alpha=D/A$
parameter expressing the ratio between the dispersion ($D$) and absorption
($A$) of the microwave radiation when it probes a metallic material.

\section{Results and Discussion}

\subsection{ESR on La$_{1-x}$Gd$_{x}$Pt$_{4}$Ge$_{12}$ }

An ESR experiment probing a small amount of paramagnetic ions in a
non-paramagnetic host lattice is supposed to bring important information
concerning not only the ionic impurity state but also the host. In
the present case, Gd is an isoelectronic substitution and is expected
to have only a local effect on the lattice parameter. 

\begin{figure}
\begin{centering}
\includegraphics[scale=0.28]{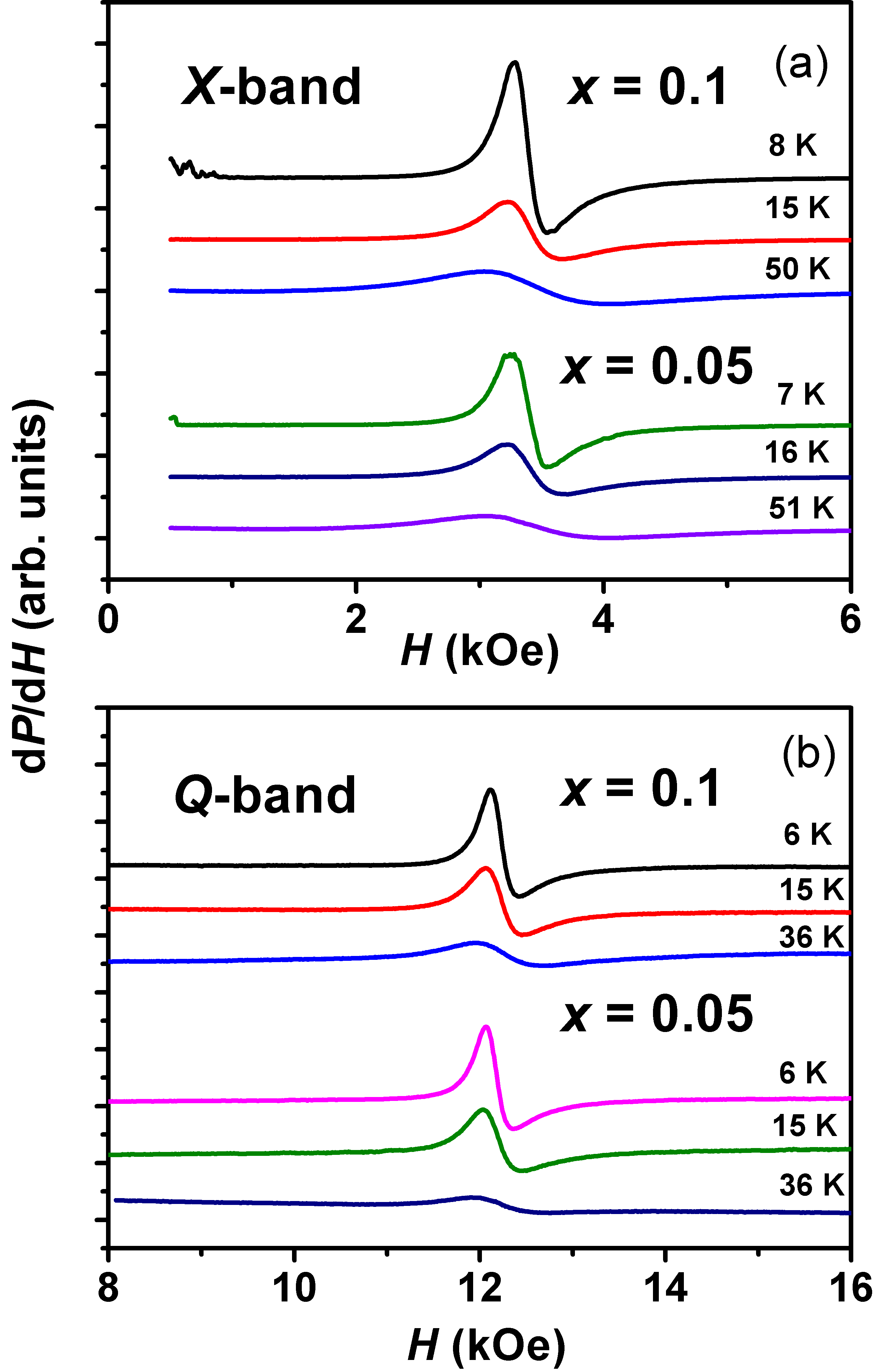}
\par\end{centering}

\caption{(Color online) Selected a) $X$-band and b) $Q$-band ESR spectra
of the derivative of the resonance absorption ($dP/dH$) for La$_{1-x}$Gd$_{x}$Pt$_{4}$Ge$_{12}$
($x=0.1$ and $x=0.05$). A single asymmetric Lorentzian lineshape
is observed in the whole temperature interval covered by the experiment.
\label{fig:Gdspectrum}}

\end{figure}

In Fig. \ref{fig:Gdspectrum} we show some selected ESR spectra taken
at both $X$-band and $Q$-band frequencies. In the whole temperature
interval, we observe a single Lorentzian lineshape and, even in the
lowest temperature of the experiment, no sign of unresolved $\mathrm{CF}$
effects were observed in the powder spectra.

\begin{figure}
\begin{centering}
\includegraphics[scale=0.27]{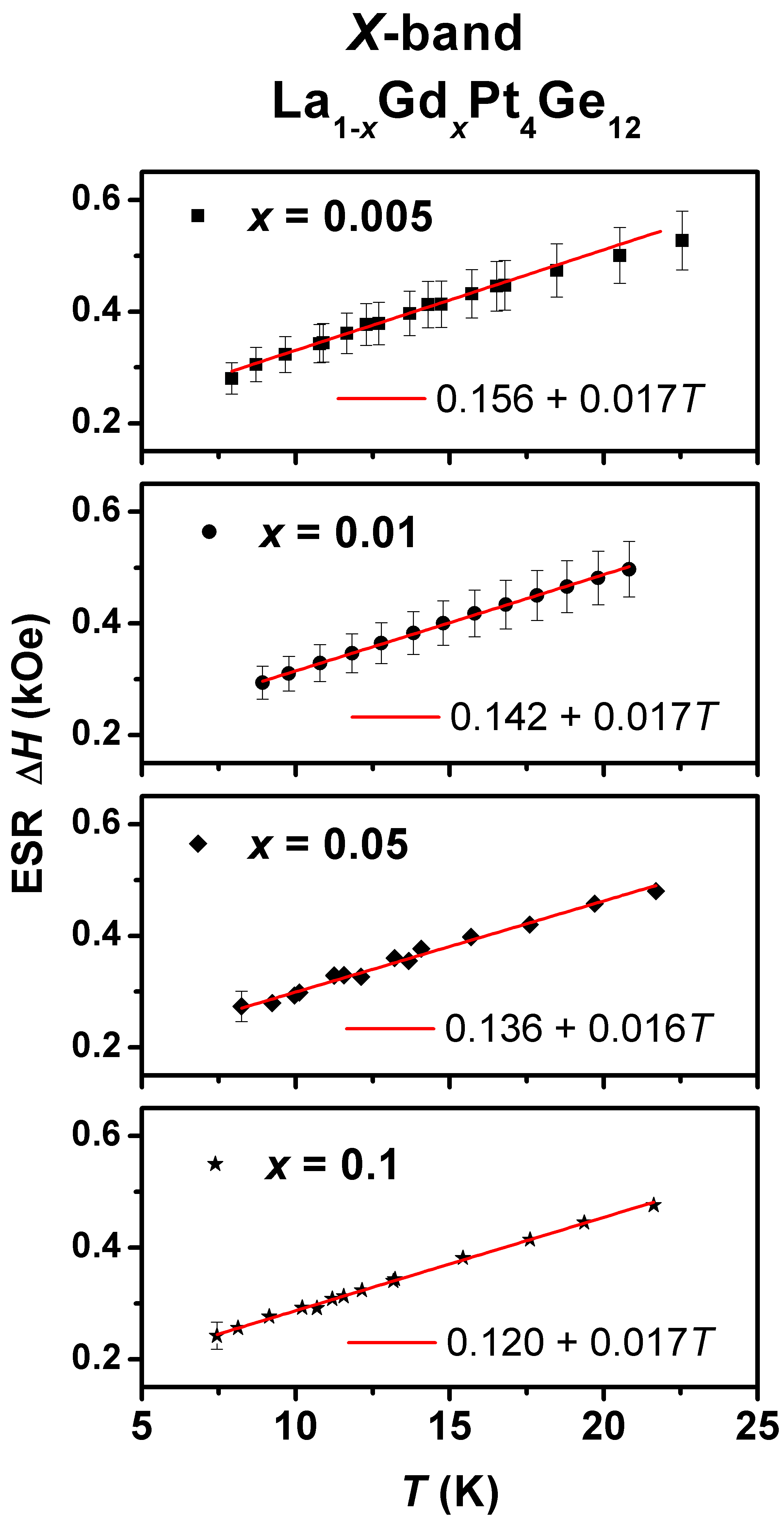}
\par\end{centering}

\caption{(Color online) Low-temperature X-Band ESR linewidth ($\Delta H$)
for different substitutions of Gd in La$_{1-x}$Gd$_{x}$Pt$_{4}$Ge$_{12}$.
The thick  line is the linear fitting to the expression $\Delta H(T)=a+bT$
and the obtained parameters are shown in the figures. \label{fig:lwXbandGd}}

\end{figure}

Fig. \ref{fig:lwXbandGd} shows the low-temperature $X$-band ESR
linewidth ($\Delta H$) for four different substitutions of Gd. It
is clear that there are no concentration effects in the order of magnitude
of $\Delta H$. In the figure, the continuous line is the fitting
to the usual Korringa expression $a+bT$. The residual linewidth $a$
is larger for lower concentrations. This quantity is expected to be
determined by CF effects, dipolar interactions and spin-spin exchange
narrowing. The latter effect renders the residual linewidth narrower
for higher concentrations and seems to be dominating over the broadening
effects of the dipolar fields. The important result, however, is that
the apparent Korringa rate $b$ is concentration independent.

In the context of the bottleneck effect, we recall that since $1/T_{\mathrm{ceS}}$
is directly proportional to the concentration of the ESR probe (Gd$^{3+}$
ions in this case), an increase of this concentration should slow
down even more the apparent relaxation rate. Thus, the hallmark of
a bottlenecked system is a concentration-dependent relaxation. The
ESR $\Delta H$ should evolve as in a Korringa-like relaxation $\Delta H(T)=a+\frac{\beta}{x}T$
, where $\beta$ is a constant, which in first approximation does
not depend on the exchange interaction, $x$ is the concentration
of the ESR probe and $a$ is the residual linewidth $\Delta H(T=0)$.
The physics of the bottleneck effect was well explored in the classical
ESR literature \citep{rettori_electron-spin_1973,rettori_dynamic_1974,davidov_electron_1973}
and also in the review by Barnes \citep{barnes_theory_1981}.

Our result indicates that the system is non-bottlenecked even for
relatively high concentrations. From our previous discussion, one
can realize that there are two possible routes for opening a bottleneck,
being either increasing $1/T_{\mathrm{ceL}}$ or lowering $1/T_{\mathrm{ceS}}$.
In LaPt$_{4}$Ge$_{12}$, the conduction electrons at the Fermi surface
originate mainly from the Ge $4p$ electrons, which are more strongly
coupled to the lattice than $s$-band electrons would be. This contributes
to the increase of $1/T_{\mathrm{ceL}}$. On top of this effect, one
should also consider that the modification of the total vibrational
dynamics of the system, as implied by the partial substitution of
La by Gd, contributes to the scattering of the conduction electrons
thus also increasing $1/T_{\mathrm{ceL}}$. 

In the non-bottlenecked regime the spin relaxation of a paramagnetic
ion in a metallic host should be described by the Korringa process.
The expected value for the Korringa rate (for a $q$-independent exchange
coupling and in the absence of multi-band effects) is given by \citep{barnes_theory_1981}:

\begin{equation}
b=\frac{\pi k_{B}}{g\mu_{B}}\Delta g^{2}\label{eq:korringa}\end{equation}

where $\Delta g$ is the so-called $g$-shift, which occurs in metals
due to the exchange interaction between the localized spins and the
conduction electrons \citep{barnes_theory_1981}. A $g$-shift is
calculated in reference to the $g$-values of experiments in insulators
($\Delta g=g_{\mathrm{exp}-}g_{\mathrm{ins}}$). Adopting the low-$T$
value of the $g$-shift, we have $\Delta g=0.009(4)$. The obtained
value for the Korringa rate ($b\approx2.3$ Oe/K) is much lower than
the one observed. Going through the relation between $b$, $\Delta g$
and the exchange interaction in more detail, we write the full expression
for both quantities \citep{barnes_theory_1981}:

\[
\Delta g=\left\langle \eta(E_{F})J(q=0)\right\rangle _{\mathrm{Av}}=\eta(E_{F})\left\langle J(q=0)\right\rangle _{\mathrm{Av}}\]

\begin{equation}
\equiv\eta(E_{F})J_{1}\label{eq:deltag}\end{equation}

\[
b=\frac{\pi k_{B}}{g\mu_{B}}(\left\langle (N(E_{F})J(k_{F},k_{F}^{'}))^{2}\right\rangle _{\mathrm{Av}})=\frac{\pi k_{B}}{g\mu_{B}}\eta(E_{F})^{2}\left\langle J(q)^{2}\right\rangle _{\mathrm{Av}}\]

\begin{equation}
\equiv\frac{\pi k_{B}}{g\mu_{B}}\eta(E_{F})^{2}J_{2}^{2}\label{eq:bkorringa}\end{equation}

where in equations \ref{eq:deltag} and \ref{eq:bkorringa} the brackets
denote an average over the Fermi surface and $\eta(E_{F})$ is the
density of states for a given spin direction at the Fermi surface
(states eV$^{-1}$mol$^{-1}$spin$^{-1}$). From these, one can see
that $\Delta g$ is a homogeneous polarization of the Gd$^{3+}$ spins
due to the exchange interaction with the conduction electrons, whereas
$b$ is related to a scattering process. Thus, in writing Eq. \ref{eq:korringa}
we have assumed $J_{1}=J_{2}$ which does not seem to be the case
in our experiment.

As extensively discussed in the literature \citep{rettori_electron-spin_1973,davidov_electron_1973,barnes_theory_1981},
this contrast, in the absence of a bottleneck, may have its origin
in multi-band and/or electron-enhancement effects. However, previous
works on the host compound do not support either the presence of different
electronic contributions to the Fermi surface or significant electronic
correlations \citep{gumeniuk_superconductivity_2008}. We shall also
discuss this issue later in this section, in connection with our $Q$-band
measurements.

A general analysis of the bottleneck phenomenon \citep{barnes_theory_1981}
shows that, if the spin-orbit scattering due to the magnetic impurity
(meaning the dependency of $1/T_{\mathrm{ceL}}$ on the amount of
Gd$^{3+}$ spins) is stronger than the effective exchange scattering
(that favors the Korringa relaxation), the system is non-bottlenecked.
Pursuing this line, we suggest that, even if the homogeneous polarization
(given by Eq. \ref{eq:deltag}) is small, the exchange scattering
process could be enhanced by the rattling modes. In this interpretation,
the rattling modes are involved in the exchange scattering process
which still dominates the relaxation. 

\begin{figure}
\begin{centering}
\includegraphics[scale=0.25]{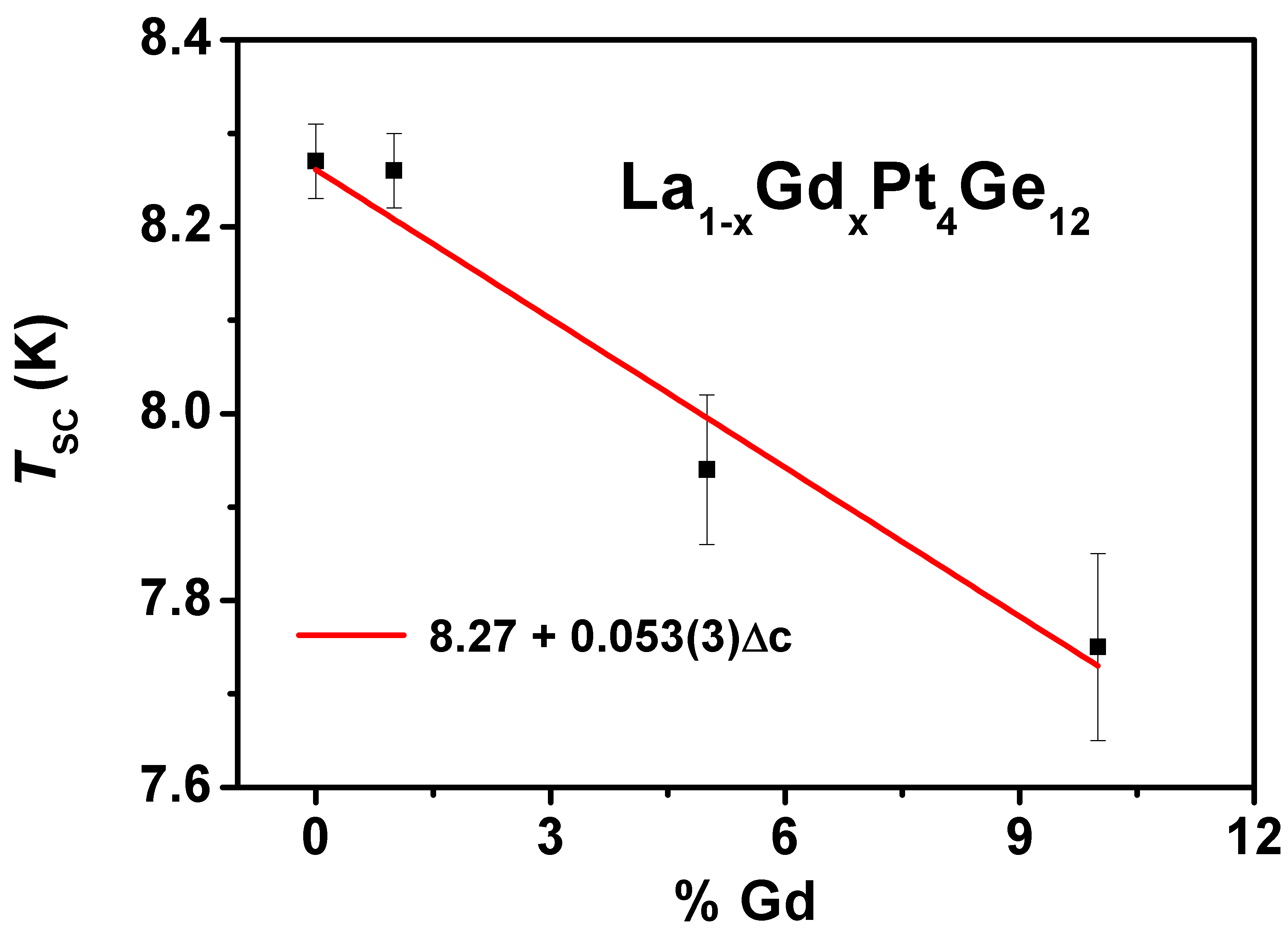}
\par\end{centering}

\caption{(Color online) Suppression of the superconducting state in La$_{1-x}$Gd$_{x}$Pt$_{4}$Ge$_{12}$
by the inclusion of Gd impurities. The thick line is the linear fit
to our experimental data. The value of $T=8.27$ K is the transition
temperature ($T_{SC}$) for the pure LaPt$_{4}$Ge$_{12}$ compound.
\label{fig:SPCsupression}}

\end{figure}

The suppression of the superconducting state ($\Delta T_{\mathrm{sc}}=T_{\mathrm{sc}}(x=0)-T(x)$)
by a given concentration ($\Delta x$) of magnetic impurities is also
given by an effective exchange scattering usually comparable with
the exchange scattering given by the ESR \citep{davidov_electron_1973}.
The process is described by the Gorkov-Abrikosov expression:

\begin{equation}
\left|\frac{\Delta T_{\mathrm{sc}}}{\Delta x}\right|=\frac{\pi^{2}}{8k_{B}}J_{\mathrm{eff}}^{2}\eta(E_{F})(g_{J}-1)^{2}J(J+1)\label{eq:abrikGork}\end{equation}

where $J_{\mathrm{eff}}$ is the effective exchange scattering, $g_{J}$
is the Landé gyromagnetic factor, and $J$ is the impurity total angular
momentum (here $J=S$, since for Gd$^{3+}$ $L=0$). In Fig. \ref{fig:SPCsupression}
we show the rather small suppression of $T_{SC}$ by the effect of
magnetic impurities. The data was obtained from magnetization measurements
measured in warming in $20$ Oe field after zero field cooling. $T_{SC}$
was defined by extrapolation of the steepest slope of $\chi(T)$ to
$\chi(T)=0$. Using the calculated density of states at the Fermi
level \citep{gumeniuk_superconductivity_2008} $\eta(E_{F})=13.4$
states eV$^{-1}$f.u.$^{-1}$$=15.8$ states eV$^{-1}$mol$^{-1}$spin$^{-1}$
in Eqs. \ref{eq:bkorringa} and \ref{eq:abrikGork}, we obtain the
following estimate for the effective exchange scattering: $J_{\mathrm{eff}}^{\mathrm{ESR}}=1.54\pm(0.40)$
meV and $J_{\mathrm{eff}}^{\mathrm{sc}}=1.24\pm(0.20)$ meV, which
agrees reasonably well with $J_{eff}^{ESR}$ and make a strong case
in favor of a relaxation process governed by the exchange interaction.
It is also a piece of evidence that the relaxation is not in a bottleneck
regime. Nevertheless, we still lack a mechanism for the enhancement
of $J_{2}=J_{\mathrm{eff}}$ in comparison with $J_{1}$ from Eq.
\ref{eq:deltag}. 

\begin{figure}
\begin{centering}
\includegraphics[scale=0.3]{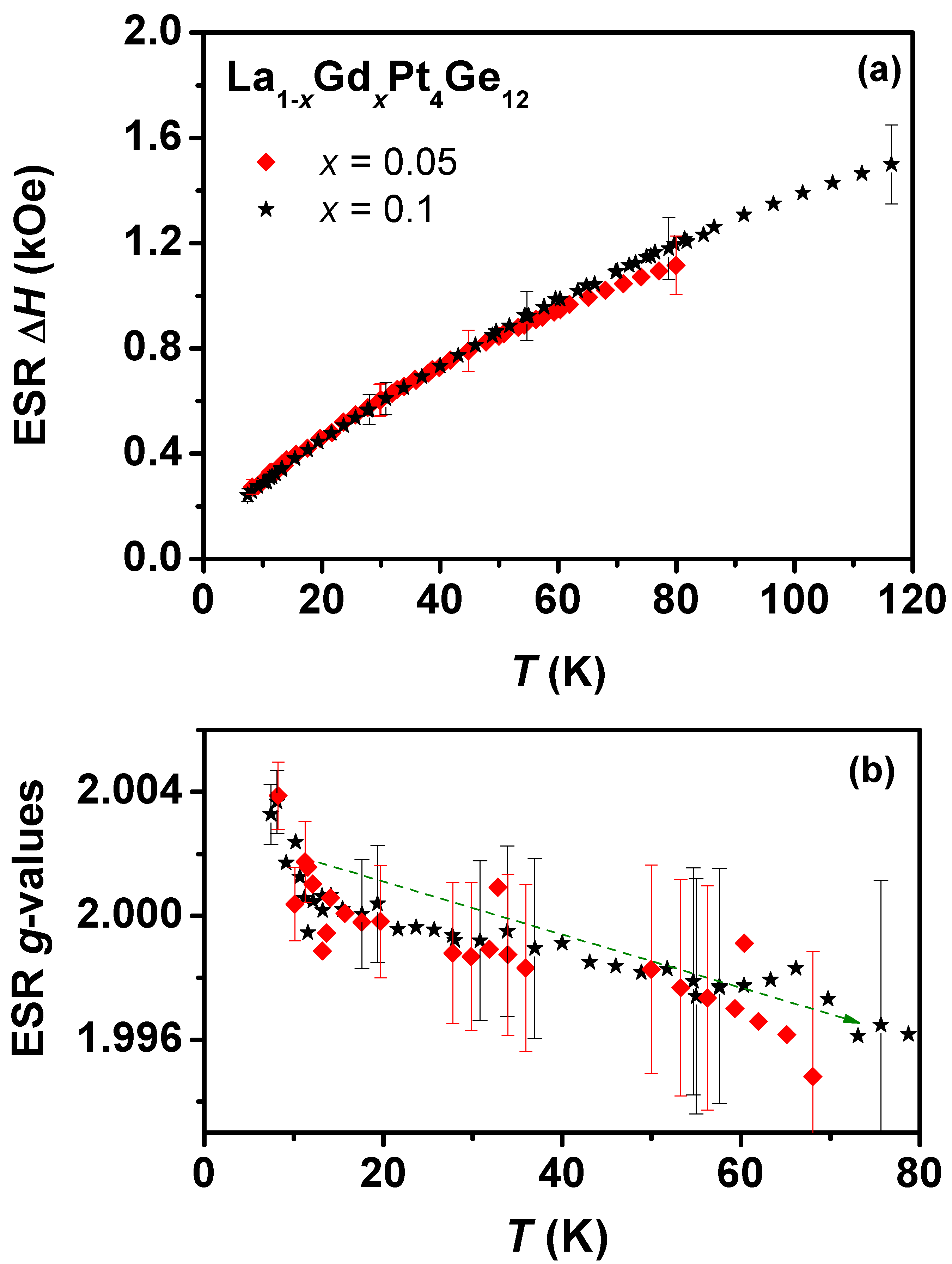}
\par\end{centering}

\caption{(Color online) (a) X band ESR linewidth ($\Delta H$) ($8\leq T\leq120$
K) and (b) $X$-band ESR $g$-values for La$_{1-x}$Gd$_{x}$Pt$_{4}$Ge$_{12}$
with $x=0.05$ and $x=0.1$. The dashed line is a guide to the eye
(see text for a discussion). \label{fig:lwXbandGdHighTgvalues}}

\end{figure}

Figure \ref{fig:lwXbandGdHighTgvalues}(a) shows the $X$-band $\Delta H$
for the two highest concentrations ($x=0.05$ and $x=0.1$) which
allows one to follow the temperature evolution of the resonance spectra
up to higher temperatures, still with relatively high resolution.
It is shown that $\Delta H$ deviates from the apparent linear behavior
at low $T$. Furthermore, the data suggest that the high temperature
relaxation regime has a weak dependence on the concentration of the
ESR probe. In previous ESR experiments in metallic cage systems, $d(\Delta H)/dT$
was also shown to be non-constant \citep{sichelschmidt_electron_2005,holanda_electron_2009}.
 In particular, it was found that the deviations from the linear behavior
may appear as a {}``kink'' in $\Delta H(T)$, which defines a low-temperature
and a high-temperature Korringa-like rate \citep{holanda_electron_2009}.

In the above cited works \citep{sichelschmidt_electron_2005,holanda_electron_2009},
it was speculated that the origin of the temperature dependence of
$d(\Delta H)/dT$ is due to some reorganization of the Fermi surface
implied by structural effects and its concomitant changes in the band
structure \citep{sichelschmidt_electron_2005,holanda_electron_2009}.
In addition, it is generally accepted that by multiple filling (filled
skutterudites with more than one element as guest) one can induce
a larger disorder in the lattice structure of the material, thus causing
a broadening in their phonon spectra \citep{peng_thermoelectric_2008,yang_dual-frequency_2007}.
In this sense, the origin of the concentration dependence of $d(\Delta H)/dT$
at high temperature can be explained by a combination of both effects:
a change in the disorder of the phonon spectra, due to different Gd
concentration, would lead to a different energy scale where the putative
reorganization of the Fermi surface takes place resulting in a weak
dependence of the high temperature relaxation on the Gd concentration.
However, we would like to point out that, as implied by Eqs. \ref{eq:deltag}
and \ref{eq:bkorringa}, similar effects would rise due to changes
in the exchange scattering.

Any effect in the Fermi surface, as well in the exchange scattering,
would also be reflected in the ESR $g$-values. Therefore, the observed
change of about $1/2$ in the value of $b$ should also manifest itself
in the $g$-values. In Fig. \ref{fig:lwXbandGdHighTgvalues}(c) we
present these data, again for the $x=0.05$ and $x=0.1$ samples.
Although the result is encouraging, one should take it carefully,
since the estimated error bars are about half the total variation.
However, our data clearly suggests a downturn in the $g$-values.

\begin{figure}
\begin{centering}
\includegraphics[scale=0.3]{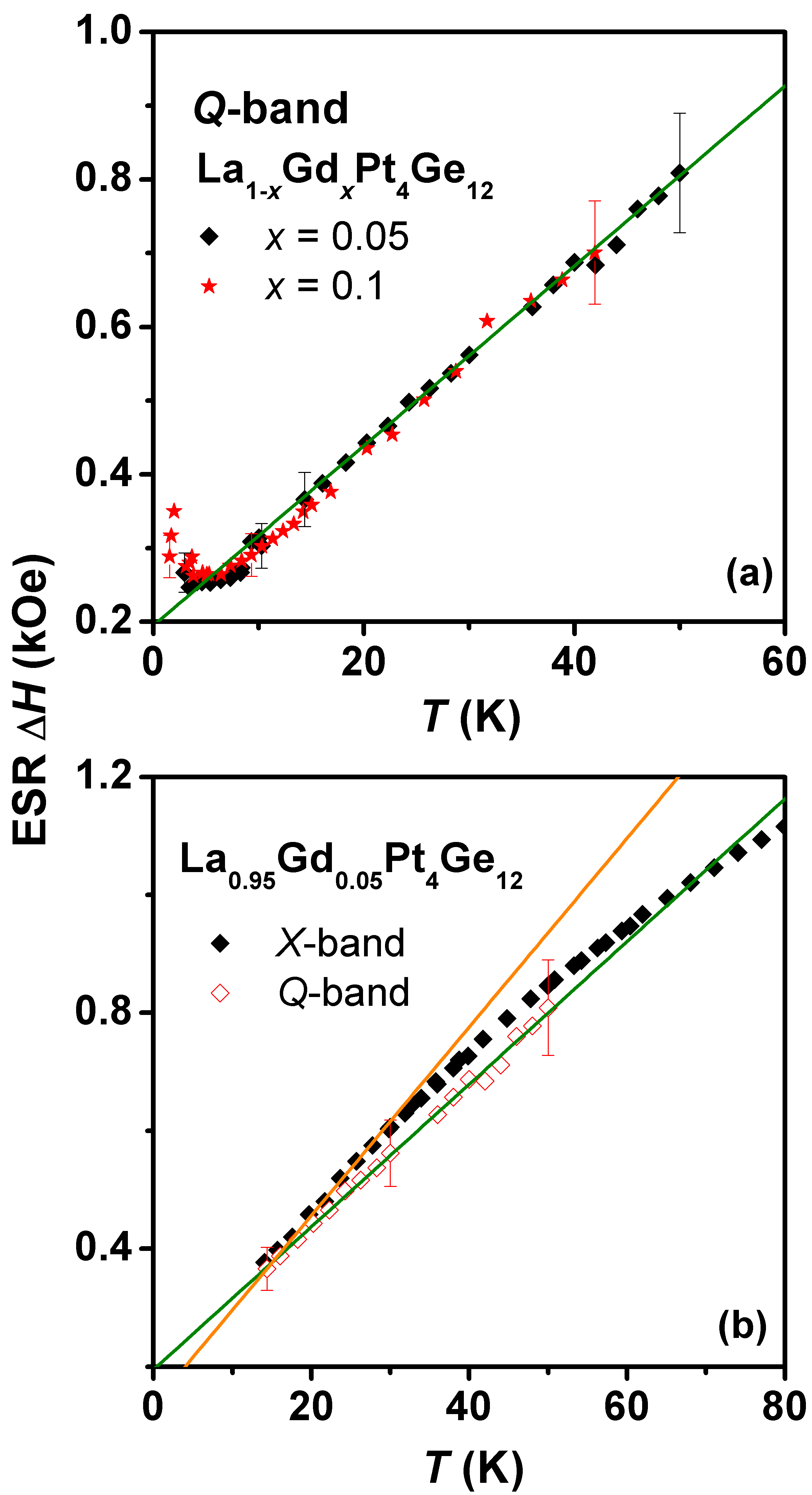}
\par\end{centering}

\caption{(Color online) a) $Q$-band ESR linewidth ($\Delta H$) for La$_{1-x}$Gd$_{x}$Pt$_{4}$Ge$_{12}$
($x=0.05$ and $x=0.1$). The thick  line is the linear fit to the
expression $\Delta H(T)=a+bT$ and the obtained parameters are $a=195(1)$
Oe and $b=12(1)$ Oe/K. Again, no concentration dependency of $\Delta H$
(apart from the low-temperature broadening) is observed. (b)\textbf{
}$Q$-band and $X$-band data for La$_{0.95}$Gd$_{0.05}$Pt$_{4}$Ge$_{12}$.
 \label{fig:lwQbandGd}}

\end{figure}

The vibrational dynamics of the guest ion may give rise to local inhomogeneities
of the crystal field which could in principle appear in an ESR experiment
as an inhomogeneous broadening of the resonance \citep{garcia_coexisting_2009}.
This may be revealed in an experiment at higher frequencies. In Fig.
\ref{fig:lwQbandGd}(a)-(b) we show our $Q$-band measurements. It
is noteworthy that no such broadening takes place, demonstrating that
the resonances are homogeneous. This means that no static disorder
is being detected in our experiment (compare the order of magnitude
of $\Delta H$ in these figures and Fig. \ref{fig:lwXbandGdHighTgvalues}
(a)-(b)). 

In Fig \ref{fig:lwQbandGd}(a) one can note that $\Delta H$ broadens
at low temperature, probably due to spin-spin interactions. Figure
\ref{fig:lwQbandGd}(a) also shows that above the temperature at which
the low-temperature magnetic fluctuations cease to dominate $\Delta H$
($T\gtrsim10$ K), the evolution of $\Delta H$ with temperature is
fairly described by a linear behavior with $b=12$ Oe/K. To highlight
this difference, we present in Fig. \ref{fig:lwQbandGd}(b) the $X$-band
and $Q$-band $\Delta H$ for the sample La$_{0.95}$Gd$_{0.05}$Pt$_{4}$Ge$_{12}$
. In the $Q$-band measurements, a systematic variation of the $g$-values
was not observed and a concentration and temperature independent $g$-value
of $g=2.011(5)$ was found. The expected Korringa rate (Eq. \ref{eq:korringa})
obtained from this result is $b=9$ Oe/K which in turn indicates,
within experimental error, that in the experiment at $Q$-band the
relaxation of the Gd$^{3+}$ ions follows the simple picture given
by Eq. \ref{eq:korringa}. Therefore, our disregard of multiband effects
in the analysis of the $X$-band data is also supported by these results,
since one cannot expect that the typical fields involved in the $Q$-band
measurements would eventually suppress any multiband effects.

Put together, these results favors the idea that, as far as ESR is
concerned, the rattling behavior manifests itself through a coupling
with the exchange interaction. A more comprehensive discussion on
this topic will be given by the end of section C. Now, we point out
that it has been recognized \citep{barnes_effect_1974} that the Ruderman-Kittel-Kasuya-Yosida
(RKKY) interaction may give rise to inhomogeneous broadening and shift
of the ESR spectra. Hence, it is plausible that an intricate relation
between the Gd$^{3+}$ rattling and the RKKY interaction between the
Gd$^{3+}$ spins, explains our results. 

We mention again that such field dependence of the relaxation was
not reported in the NMR experiments \citep{nakai_low-lying_2008,kanetake_superconducting_2010}.
We also stress that the absence of static inhomogeneities is an important
sign of the sample quality, in particular, it demonstrates that the
Gd ions are homogeneously distributed in the sample.

\subsection{ESR on La$_{1-x}$Eu$_{x}$Pt$_{4}$Ge$_{12}$ and EuPt$_{4}$Ge$_{12}$ }

The partial substitution of La for Eu adds one hole to the Fermi level
per f.u.. Thus, it is expected that this local electronic inhomogeneity
gives rise to interesting effects in the interaction of the Eu$^{2+}$
spin with the conduction electrons. 

\begin{figure}
\begin{centering}
\includegraphics[scale=0.27]{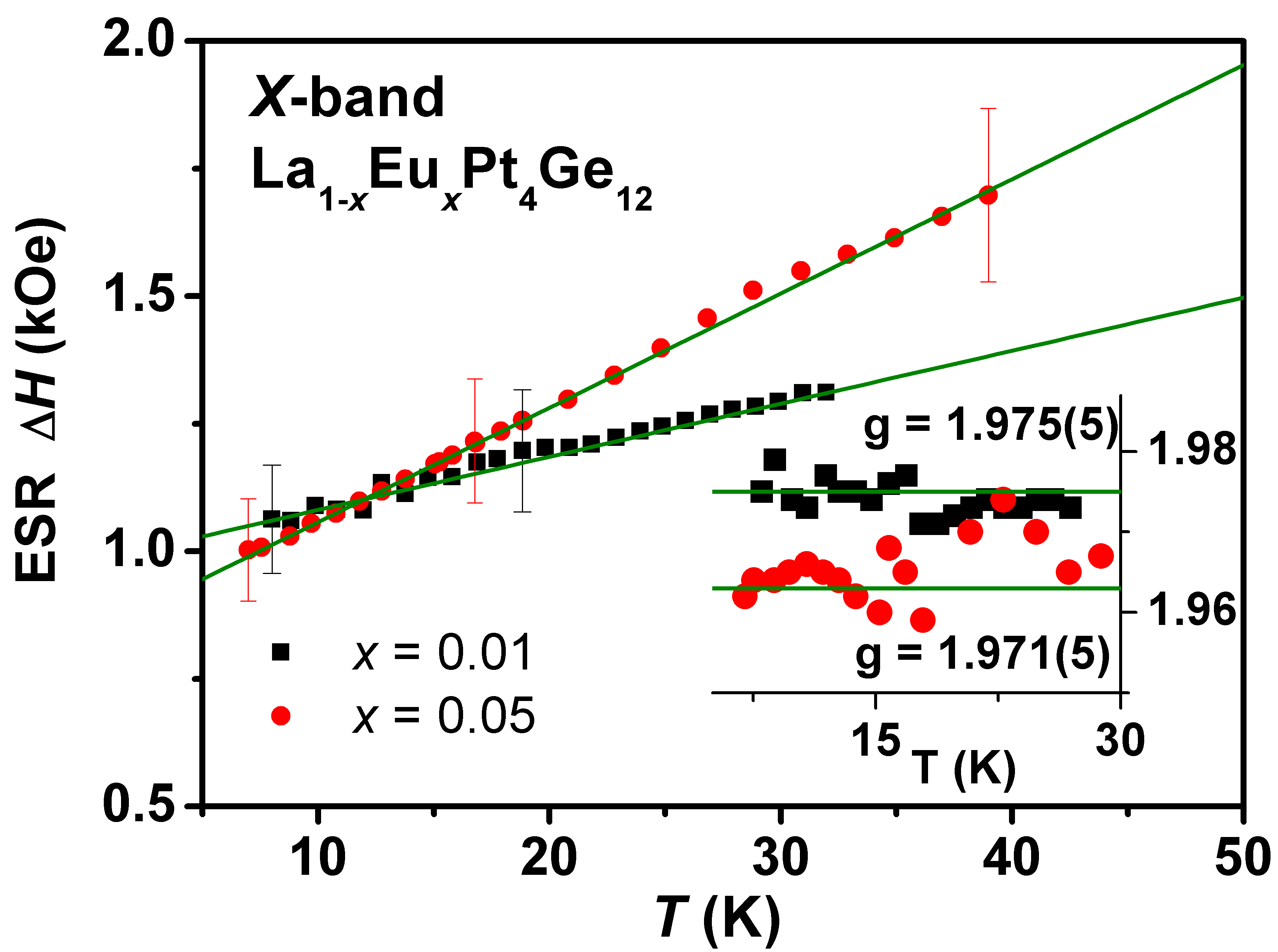}
\par\end{centering}

\caption{(Color online) $X$-band ESR linewidth ($\Delta H$) for La$_{1-x}$Eu$_{x}$Pt$_{4}$Ge$_{12}$
($x=0.01$ and $x=0.05$). The thick line is the linear fit to the
expression $\Delta H(T)=a+bT$. For $x=0.01$ the obtained coefficients
are $a=977$ Oe and $b=10.5(5)$ Oe/K whereas in the case $x=0.05$
the coefficients are $a=834(5)$ Oe and $b=22.5(3)$ Oe/K. The inset
shows a temperature independent $g$-value of about $g=1.975(5)$,
for $x=0.01$, and $g=1.963(5)$ for $x=0.05$, implying in a negative
$g$-shift (see text for discussion). \label{fig:lwXbandEu}}

\end{figure}

We present in Fig. \ref{fig:lwXbandEu} the $X$-band ESR $\Delta H$
for La$_{1-x}$Eu$_{x}$Pt$_{4}$Ge$_{12}$ ($x=0.01$ and $x=0.05$).
Again, no signs of a bottleneck effect were found. In comparison with
our previous discussion, the huge residual linewidth ($a\approx1000$
Oe) is most likely due to relatively larger CF effects \citep{barnes_theory_1981}.
In the inset of Fig. \ref{fig:lwXbandEu}, the thick solid lines represent
the average of the measured $g$-values for each sample. From these
values, one obtains relatively large $\Delta g=-0.018(5)$ and $\Delta g=-0.030(5)$
for the $x=0.01$ and $x=0.05$ samples, respectively. These $g$-shifts
(c.f Eq. \ref{eq:korringa}) give a Korringa rate of $b=9(3)$ Oe/K
and $b=28(8)$ Oe/K, respectively. In the former case, the measured
Korringa rate is $b=10.4(5)$ Oe/K and in the latter case $b=22.4(5)$
Oe/K. Therefore, a comparison of the results suggests that for both
$x=0.01$ and $x=0.05$, the Eu$^{2+}$ relaxation fits in the simple
picture of Eq. \ref{eq:korringa}. 

A negative $\Delta g$ originates from a covalent mixing between localized
and itinerant states \citep{barnes_theory_1981,schrieffer_relation_1966}.
In the case of 4$f$ impurities, a negative $\Delta g$ results from
the mixing of 4$f$ states and $d$ itinerant electrons. Thus, our
results, for both the Korringa rates and $g$-shifts, ask for a significant
change of the local electron density at the Eu impurities, even for
small substitution of La by Eu. 

\begin{figure}
\begin{centering}
\includegraphics[scale=0.3]{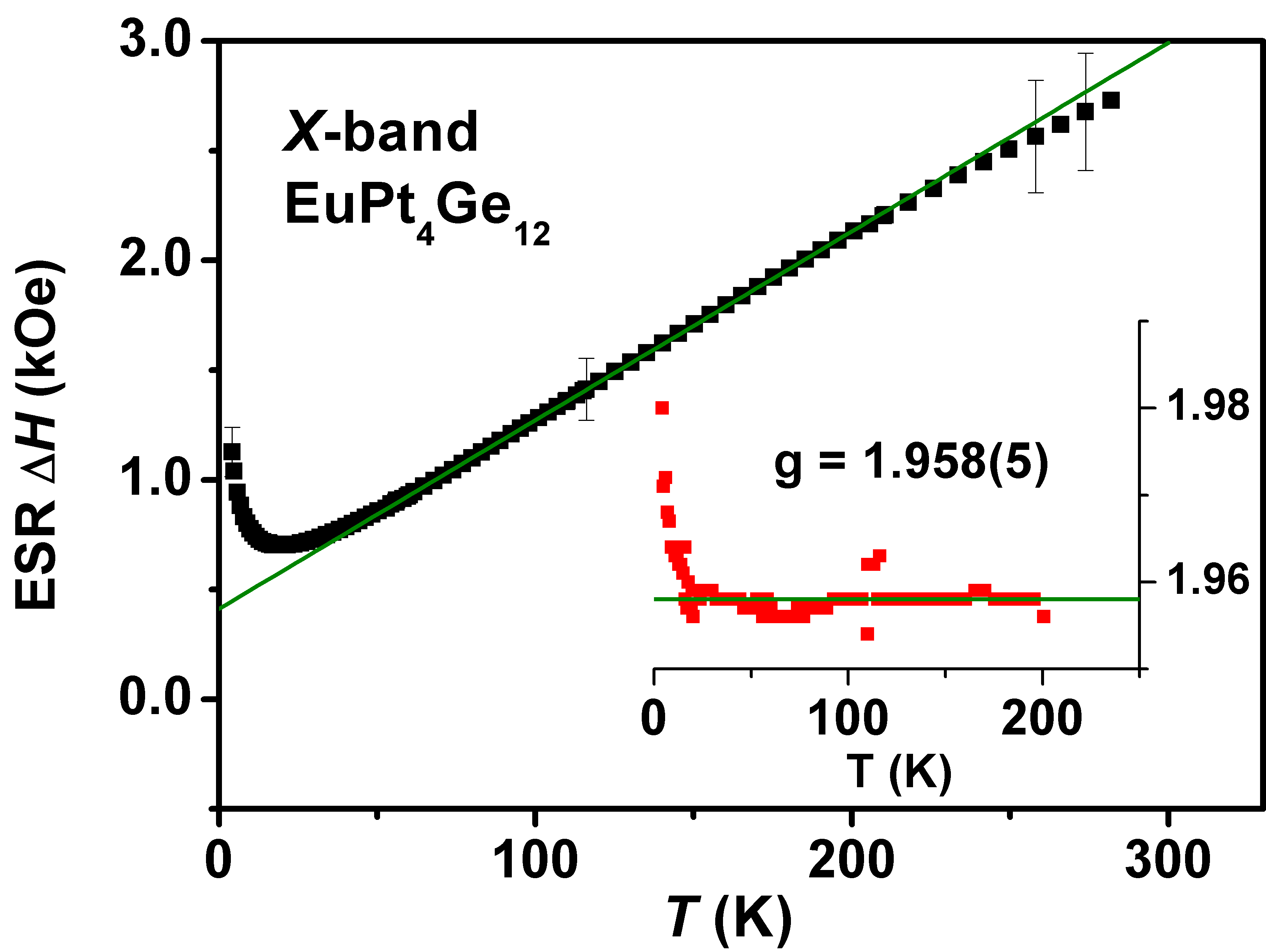}
\par\end{centering}

\caption{(Color online) $X$-band ESR linewidth ($\Delta H$) for EuPt$_{4}$Ge$_{12}$
. The thick line is the best fit to the expression $\Delta H(T)=a+bT$,
and the obtained coefficients are $a=412(2)$ Oe and $b=8.6(1)$ Oe/K.
The line describes the behavior of $\Delta H$ for $T\geq50$ K. In
the inset, the ESR $g$-values are presented. For $T\geq20$ K, $g=1.958(5)$
is observed, implying in a negative $g$-shift. Below this temperature
the $g$-values increase (see text for discussion). \label{fig:lwXbandEuPt4}}

\end{figure}

Different from the Gd case, the ternary EuPt$_{4}$Ge$_{12}$ compound
was available for studies. Figure \ref{fig:lwXbandEuPt4} presents
our results. Here we find $\Delta g=-0.035(5)$ (for $T\gtrsim20$
K), implying, by Eq. \ref{eq:korringa}, a Korringa rate $b\approx34$
Oe/K. The measured Korringa rate for the concentrated EuPt$_{4}$Ge$_{12}$
system is $b=8.7(3)$ Oe/K. This enormous contrast is usually attributed
to the $q$-dependency of the exchange coupling $J_{\mathrm{eff}}$,
and provides evidence for an indirect exchange interaction between
the localized spins of the RKKY type \citep{taylor_electron_1975}. 

Alternatively, it mighty be  that for the concentrated system ($x=1$)
the bottleneck regime is eventually reached, resulting in a slowing
down of the apparent relaxation rate. However, in favor of the presence
of an RKKY interaction, we point out that the negative $\Delta g$
found in this case ($\Delta g=-0.035(5)$, $x=1$) cannot be due to
changes in the local electron density, but should reflect an important
distinction between the Fermi surfaces of LaPt$_{4}$Ge$_{12}$ and
EuPt$_{4}$Ge$_{12}$. Hence, since the presence of $d$ electrons
favors the spin-orbit scattering (and thus $1/T_{\mathrm{ceL}}$),
the system should not be bottlenecked or, at least, not severely bottlenecked.
As an important consequence, the evolution of $\Delta H$ is still
due to the exchange interaction between the local and itinerant spins
, which is now $q$-dependent. 

Figure \ref{fig:lwXbandEuPt4} also reveals a low-temperature increase
of $\Delta H$. EuPt$_{4}$Ge$_{12}$ is known to order magnetically
below $1.78$ K. Specific heat and dc-susceptibility display four
anomalies which indicate a complex magnetic phase diagram \citep{nicklas_magnetic_2011}.
Due to the sensitivity of the ESR technique, it is possible that the
broadening of $\Delta H$ is related to the magnetic fluctuations
which give rise to these different ordered states. The inset shows
that the $g$-value increases for $T\lesssim20$ K, most likely due
to the onset of the internal fields associated with the discussed
magnetic phases at low temperatures. 

In the whole temperature interval investigated, no indications could
be found for an Eu$^{2+}$ rattling behavior. The results of our $Q$-band
investigation are very similar and do not add to our discussion.

\subsection{Possible role of a Raman relaxation}

Following the discussion of Dahm \emph{et al.} \citep{dahm_nmr_2007},
it is tempting to look for other sources for the relaxation of the
localized ions, specially in the case of the Gd-substituted samples.
It is possible that the substitution of even a small amount of La
by Gd, gives rise to a significant change in the guest vibrational
dynamics. In this context it has to be mentioned that a ternary phase
GdPt$_{4}$Ge$_{12}$ does not exist as a stable phase under ambient
conditions. Indeed, some threshold requirements for the filled atoms
in order to stabilize a skutterudite have to be fulfilled. Several
studies have been devoted to probe high-pressure synthesis routes
to compounds that are based on cations with radii below the critical
values (see \citep{gumeniuk_filled_2010} and references therein).
However, attempts to synthesize GdPt$_{4}$Ge$_{12}$ by similar methods
failed. Instead, phases with different crystal structures form.

The theory of Dahm \emph{et al.} was written in terms of NMR parameters
and does not allow to make a direct comparison with our results. Nevertheless,
given that the methods for the study of the relaxation processes are
quite general, we expect that the qualitative aspects of the theory
will remain valid. Therefore, we simply add a scaling parameter to
match the theoretical curve with our results.

This theory considers that the guest vibrational dynamics may be described
by a Hamiltonian for the anharmonic oscillator, which is treated in
a quasi-harmonic approximation. This treatment gives rise to a non-linear
equation for an effective frequency which, now, is temperature dependent:

\begin{equation}
\left(\frac{\omega_{0}}{\omega_{00}}\right)^{2}=1+\alpha\frac{\omega_{00}}{\omega_{0}}\left(\frac{1}{\exp(\hbar\omega_{0}/k_{B}T)-1}-\frac{1}{2}+\frac{1}{2}\frac{\omega_{0}}{\omega_{00}}\right)\label{eq:nonlinear}\end{equation}

where $\omega_{0}$ is the effective phonon frequency, $\alpha>0$
is a dimensionless parameter characterizing the amount of anharmonicity
and $\omega_{00}=\omega_{0}$ at $T=0$. In the simplest rattling
approximation, $\omega_{0}=\omega_{00}=\theta_{E}$. The relaxation
rate $1/T_{1}T$ which emerges from this picture gives a fair description
of the relaxation phenomena in some skutterudites \citep{nakai_low-lying_2008,kanetake_superconducting_2010}.
It puts forward the idea that the rattling modes have a fundamental
role in the spin relaxation of cage systems by inducing an indirect
phonon relaxation, usually found in experiments with insulators. 

\begin{figure}
\begin{centering}
\includegraphics[scale=0.3]{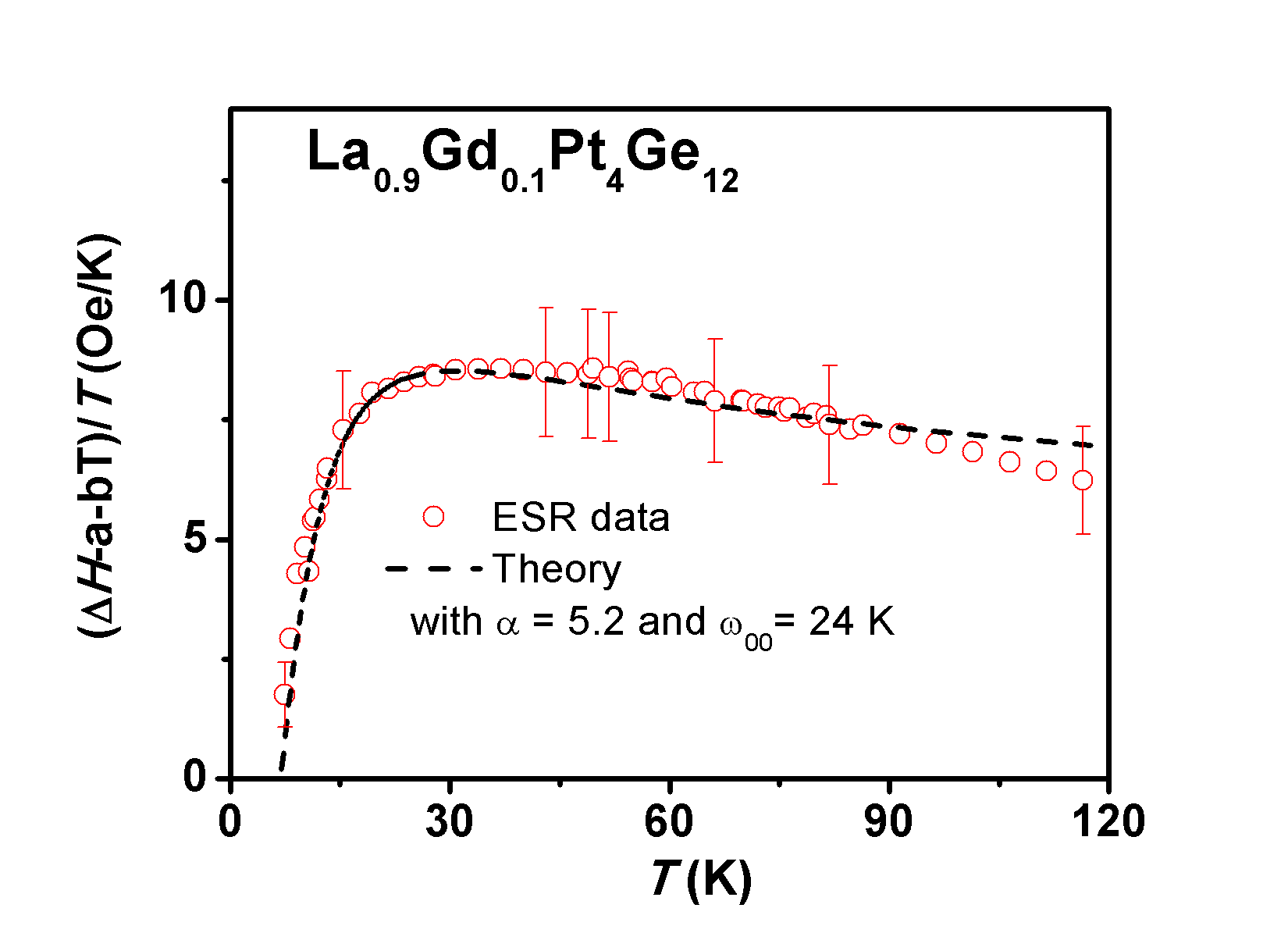}
\par\end{centering}

\caption{$X$-band $(\Delta H(T)-a-bT)/T$ for La$_{0.9}$Gd$_{0.1}$Pt$_{4}$Ge$_{12}$,
where $a$ and $b$ are the parameters determined in the $Q$-band
measurements. The resulting data are compared with the solid line
which originates from a theory for spin relaxation due to anharmonic
phonons \citep{dahm_nmr_2007}. Parameters that we used (see text)
in the calculation are shown in the figure. \label{fig:theory}}

\end{figure}

In Fig. \ref{fig:theory} we compare our experimental results with
the above presented theory. In the NMR experiments \citep{nakai_low-lying_2008,kanetake_superconducting_2010},
the contribution ascribed to the rattling modes was isolated after
a comparison between the relaxation rates of the nuclear spins of
the elements in the cage structure, and that of the rattler element.
However, since the lattice dynamics of filled skutterudites is determined
by significant hybridization of vibrational states of the guest-atoms
and of the host structure \citep{koza_vibrational_2010,koza_vibrational_2011},
such a partitioning into independent contributions is not straightforward
and therefore should be only a first approximation. Here, even such
an approximate {}``two-site'' investigation is not possible. Our
choice is to subtract the Korringa rate obtained at $Q$-band from
our linewidth data taken at $X$-band. We use the results in the case
of La$_{0.9}$Gd$_{0.1}$Pt$_{4}$Ge$_{12}$, which are obtained in
a broader temperature range. In doing that, we assume that the rattling
effects in the relaxation are quenched at $Q$-band and that at $X$-band
one may find both contributions. The value of $\omega_{00}=24$ K
found in this way is considerably lower than $\theta_{E}=96$ K obtained
from the specific heat of LaPt$_{4}$Ge$_{12}$ \citep{gumeniuk_superconductivity_2008}.
This result should be interpreted as an extra Einstein mode associated
with the Gd$^{3+}$ site, since it is exclusively the Gd$^{3+}$ relaxation
which is being probed. It is also expected that, since Gd$^{3+}$
is heavier and significantly smaller than La, it would vibrate with
a lower frequency. We also used $\alpha=5.2$, which is larger than
previously reported in the NMR experiments \citep{nakai_low-lying_2008,kanetake_superconducting_2010},
reflecting a stronger anharmonicity due to the inclusion of the Gd$^{3+}$
ions in the oversized cage. We can speculate that the effects of a
fully harmonic vibrational dynamics are averaged out and cannot be
observed in an ESR experiment, as in the previous section where we
addressed the results for the Eu$^{2+}$ ESR. Recent heat capacity
measurements, taken at high fields to suppress the superconducting
phase, reveal a phonon contribution compatible with an Einstein phonon
with $\theta_{E}\approx20$ K, in good agreement with our results
from ESR \citep{Unpub}. 

It is important to make an attempt to reconcile the analysis of sections
A and C. Dahm e\emph{t al }\citep{dahm_nmr_2007} considered that
fluctuations in the electron field gradient implied by the rattling
behavior provide the mechanism coupling the rattling modes to the
relaxation process. In the present case, it is known that the ESR
relaxation of a system of spins coupled by an interaction of the RKKY
type requires the evaluation of the dynamical part of the RKKY interaction
\citep{barnes_effect_1974}. In this case, a molecular field approximation
for the exchange coupling among the spin of the local system should
be applied. This molecular field takes into account the finite lattice
spacing between the Gd atoms and is determined by the lattice average
of the RKKY interaction. Hence, the relaxation of the spin system
will be quite sensitive to the lattice parameters. In this direction,
we can formulate a scenario where the change of the total vibrational
dynamics of the system, as implied by the dynamical behavior (rattling)
of the Gd$^{3+}$ ions, generate local fluctuations of the RKKY interaction
that are quenched at higher fields.  It explains why the evolution
of $\Delta H$ can be explored with a theory based on a Raman process
and is, nevertheless, field dependent. A similar explanation should
hold for the ESR on EuM$_{4}$Sb$_{12}$ (M$=$Fe, Ru, Os), where
the evolution of $\Delta H$ was also shown to be field dependent
in a similar way.

\section{Conclusions}

In La$_{1-x}$Gd$_{x}$Pt$_{4}$Ge$_{12}$, the relaxation process
is non-bottlenecked even for relatively high Gd concentrations. As
a consequence, the relaxation process should be determined by the
exchange coupling between the local and itinerant spin systems. Besides
this result, the effective exchange scattering estimated from the
ESR measurements (a microscopic probe), agrees well with the one estimated
from the Abrikosov-Gorkov expression for the suppression of $T_{\mathrm{sc}}$
(as determined from the total magnetization, a macroscopic probe),
due to the addition of magnetic ions. 

In addition, in $X$-band it was found that the order of magnitude
of the measured $g$-shifts implies a much smaller relaxation rate
than the one observed and it was proposed that the guest vibrational
dynamics could give rise to either an enhanced exchange scattering
rate or to an additional relaxation process. 

The field dependence of the Gd$^{3+}$ ESR relaxation suggests that
fluctuations of the RKKY local field, implied by the guest vibrational
dynamics, should play a fundamental role in the ESR response of these
systems. These were shown to be small fluctuations that are quenched
in going from $X$-band to $Q$-band measurements, at which one observes
a relaxation process as expected in the simplest Korringa picture.
An extra phonon mode, to be associated with the vibrational dynamics
of the Gd$^{3+}$ ions, was isolated in the La$_{0.9}$Gd$_{0.1}$Pt$_{4}$Ge$_{12}$
with $\theta_{E}\approx24$ K.

For the Eu-substituted samples, no signatures for rattling behavior
were found. The local lattice structure of the Gd$^{3+}$ ions in
LaPt$_{4}$Ge$_{12}$ should be more disordered, due to significant
difference in the atoms radii, than that of the Eu$^{2+}$ ions. In
this sense, this work provides a microscopic insight on how the local
vibrational spectra may change upon small replacements in the guest
site.

The evolution of $\Delta H$ with $T$ in La$_{1-x}$Eu$_{x}$Pt$_{4}$Ge$_{12}$
is well explained within the simplest approximation of a single-band
metal and $q$-independent $J_{\mathrm{eff}}$ for small contents
of Eu. The presence of the $d$ electrons at the Fermi surface is
suggested by the measured negative $g$-shift. With increasing content
of Eu, larger shifts of the resonance were observed, pointing to a
continuous change of the Fermi surface upon Eu substitution. In the
case of EuPt$_{4}$Ge$_{12}$, evidence supporting the Eu$^{2+}$
ions being coupled by the RKKY interaction was found.

In particular, our work adds another piece of evidence for the role
of the guest-ion vibrational dynamics in the spin relaxation in metallic
compounds, where the Korringa process is expected to dominate. 


\end{document}